\documentstyle[12pt]{article} \textwidth15.0cm \textheight21.0cm
\newcommand{\be}{\begin{equation}} \newcommand{\ee}{\end{equation}}
\newcommand{\bea}{\begin{eqnarray}} \newcommand{\eea}{\end{eqnarray}} \date{}

\begin{document} \begin{titlepage} \begin{flushright} HD--THEP--92--49
\end{flushright} \vspace{1.8cm} \begin{center} {\bf\LARGE Flow Equations}\\
{\bf\LARGE for the Higgs Top System}\\ \vspace{1cm} Ulrich
Ellwanger\footnote{Supported by  a DFG Heisenberg fellowship} and  Lautaro
Vergara\footnote{Supported by a DAAD fellowship}\\ \vspace{.5cm} Institut
f\"ur Theoretische Physik\\ Universit\"at Heidelberg\\ Philosophenweg 16,
D-6900 Heidelberg, FRG\\ \vspace{3cm} {\bf Abstract:}\\
\parbox[t]{\textwidth}{ The flow equations or exact RG equations for the Higgs
Top System are solved to leading order in $1/N_c$. This allows to relate
arbitrary bare actions  with this field content continuously to effective low
energy theories, and we find the flow converging towards general
renormalizable models. The  assumption of a bare action of the generalized
Nambu-Jona-Lasinio type does  not restrict the parameters of the low energy
theory. } \end{center}\end{titlepage} \newpage \setcounter{page}{2}

\section{Introduction} The experimental lower bound on the top quark mass has
revived speculations about dynamical electroweak symmetry breaking in the form
of a  top quark condensate \cite{1}-\cite{3}. A large Higgs-top Yukawa coupling
might be a residual effect of compositeness in the corresponding channel, i.e.
the Higgs field of the standard model could reveal itself as a $t\bar t$  bound
state. A concrete dynamical model can be formulated \cite{3} by assuming the
existence of non-renormalizable four Fermi interactions as in the
Nambu-Jona-Lasinio (NJL) model \cite{4} and rely on an analysis to leading
order in $1/N_c$, where $N_c$ is the number of colours. The effective low
energy theory of such a model has the form of the Higgs top sector of the
standard model with restricted parameters such that predictions on the
physical Higgs and top quark masses can be made \cite{3}. Subsequently it was
observed that these  constraints are relaxed if the original model is
generalized \cite{5}-\cite{7}. Whereas the original model contains just a
single pointlike four Fermi interaction (and a specified form of the cutoff),
the generalizations include additional irrelevant operators in the form  of
four Fermi interactions involving derivatives and general cut-off  procedures.

The underlying problem is the typical question in  quantum field theory of the
relation between a ``bare'' high energy action and an effective low energy
theory. Here the bare action does not necessarily have to be of the
renormalizable type, but could again be the result of technicolour, grand
unification, supergravity or even string theory. Nevertheless the effective low
energy action will always be of the renormalizable type in the sense that the
coefficients of all irrelevant operators (or all one-particle-irreducible Green
functions with dimensions larger than four) are computable, in perturbation
theory, in terms of the finite number of coefficients of marginal or relevant
operators (or renormalizable couplings).

An intuitive picture of the relation between the bare high energy action and
the effective low energy theory is provided by a scale-dependent effective
action $S_{int}(\mu)$, which is obtained by integrating out all physical
degrees of freedom with momenta $p$ with  $|p|>\mu$, and which interpolates
continuously between the high energy and low energy action. The scale
dependence of $S_{int}$ can be  described by means of exact renormalization
group or flow equations \cite{8}, \cite{9}, which have been introduced to
continuum quantum field theory by Polchinski \cite{10}.

These flow equations describe the RG flow of relevant, marginal, and irrelevant
couplings simultaneously and are of particularly simple form, albeit they
constitute an infinite set of coupled differential  equations. The equations
can be useful for proofs of renormalizability \cite{11}-\cite{14},  and can be
obtained within the context of an average action  \cite{15}. Recently it was
proposed to apply these equations to the formation of  bound states and
dynamical symmetry breaking by combining them with the introduction of
collective fields \cite{16}, and the generalized NJL models are an appropriate
testing ground for this formalism. In the context of a composite Higgs field
within the Standard Model, the use of flow equations has also been advertised
(but not exploited) in \cite{17}. Below we will present the solution of the
flow equations, to leading order in $1/N_c$, for the Higgs top system. Since it
allows to relate effective high and low energy theories in an explicit and
continuous way, it can be used both as  a new method to compute the effective
low energy theory, but also to gain insight  into the physics of this relation.
In particular the convergence of the RG flow, starting with an infinite number
of nonrenormalizable couplings (or irrelevant operators) at high scales,
towards an effective low energy theory  parametrized  by only a finite number
of renormalizable couplings can be studied explicitly. This way we can discuss
whether the assumption of a  bare action of the generalized NJL type restricts
the parameters of the low energy theory. The corresponding negative answer is
not new \cite{5}-\cite{7}, but in any case every exactly solvable system as the
present one constitutes a useful terrain for testing approximations to these
flow equations, which have to be developed in the future if they want to be
applied to more complicated theories as non-abelian gauge theories.

In the next section we will present the flow equations for the Higgs top system
to leading order in $1/N_c$, their solutions and implications for the effective
low energy theory. We concentrate mainly on the effective potential, whose
scale dependence within different models has also recently been studied in
\cite{18}. Section 3 is devoted to generalized NJL models, which correspond to
just a particular set of boundary conditions of the flow equations. Finally we
close with discussions and an outlook in section 4.

\section{The Flow Equations} \setcounter{equation}{0} Let us discuss, for
simplicity, the general features of the flow equations in the context of a
model with a single scalar field $\phi$, and  we will work in Euclidean space.
The flow equations rely on a cutoff propagator $P_\Lambda^{\Lambda_0}(p^2)$,
which is defined by
 \be \label{2.1}P_\Lambda^{\Lambda_0}(p^2)=\frac{f(\Lambda,p)}{p^2+m^2}\ee with
\be \label{2.2}
f(\Lambda,p)=\tilde\theta(\Lambda_0^2/p^2)-\tilde\theta(\Lambda^2/p^2).\ee
$\tilde\theta$ denotes a smeared $\infty$-differentiable $\theta$-like function
\be\label{2.3}  \tilde\theta(\Lambda^2/p^2)=\cases{ 1,& $p^2<\Lambda^2/2$\cr
{\rm smooth}& $\Lambda^2/2<p^2<2\Lambda^2$\cr 0,& $2\Lambda^2<p^2$.\cr}\ee
Hence $P_\Lambda^{\Lambda_0}(p^2)$ is non-vanishing for
$\Lambda^2/2<p^2<2\Lambda_0^2$.

Let us define $S_{int}(\varphi,\Lambda)$ by
 \be\label{2.4}
e^{-S_{int}(\varphi,\Lambda)}=e^{D_\Lambda^{\Lambda_0}}e^{-S_{int}
(\varphi,\Lambda_0)}\ee with
  \be\label{2.5} D_\Lambda^{\Lambda_0}=\frac{1}{2}
\int\frac{d^4p}{(2\pi)^4}P_\Lambda^{\Lambda_0}(p^2)\frac{\delta}
{\delta\varphi(p)}
\frac{\delta}{\delta\varphi(-p)}.\ee $S_{int}(\varphi,\Lambda)$ consists of all
Feynman diagrams generated by the interactions present in
$S_{int}(\varphi,\Lambda_0)$, which is to be identified with the bare action of
the theory. The internal propagators, however, involve both an ultraviolet
cutoff $\Lambda_0$ and an infrared cutoff $\Lambda$. It can be shown that
$S_{int}(\varphi,\Lambda=0)$ is related to the generating functional $G(J)$ of
connected Green functions of the theory \cite{13}, \cite{16}:
 \be\label{2.6} S_{int}(\varphi,0)\Bigr\vert_{\varphi=P_0^{\Lambda_0}J}=
G(J)-\frac{1}{2}\int\frac{d^4p}{(2\pi)^4}J(p)P^{\Lambda_0}_0(p^2)J (-p).\ee
Furthermore $S_{int}(\varphi,\Lambda)$ satisfies the flow equations
\cite{10}-\cite{16} \begin{eqnarray}\label{2.7} \partial_\Lambda
S_{int}(\varphi,\Lambda)&=&\frac{1}{2}\int\frac{d^4p}
{(2\pi)^4}\partial_\Lambda
P^{\Lambda_0}_\Lambda(p^2)\cdot\Bigl\lbrace\frac{\delta
^2S_{int}(\varphi,\Lambda)}{\delta\varphi(p)\delta\varphi(-p)} %\nonumber\\
-\frac{\delta S_{int}(\varphi,\Lambda)}{\delta\varphi(p)}\frac{\delta
S_{int}(\varphi,\Lambda)}{\delta\varphi(-p)}\Bigr\rbrace.\nonumber\\
{}~\end{eqnarray}

In principle the integration of the flow equations (\ref{2.7}) with specified
boundary conditions at $\Lambda=\Lambda_0$ allows the computation of
$S_{int}(\varphi,0)$, which unites the information about all physical Green
functions according to (\ref{2.6}). Generally, however,
$S_{int}(\varphi,\Lambda)$ contains apart from the relevant and marginal
operators all irrelevant operators allowed by the symmetries of the  theory.
After a decomposition of eq. (\ref{2.7}) into this infinite set of operators
one thus obtains an infinite set of flow equations for the associated
couplings. In contrast to standard $\beta$ functions, at least, the right-hand
side of these flow equations is at most quadratic in the couplings.
Nevertheless we will see  that for the Higgs top system the flow equations can
be solved exactly to leading order in $1/N_c$.

The field content of the Higgs top system is given by a four-component Dirac
spinor $\psi_a$ and a complex scalar $\phi=\sigma+i\pi$ with $\sigma$ and $\pi$
real. $a$ denotes the colour index, hence $a=1,...,N$ with $N=N_c=3$ for QCD.
The right-handed components of $\psi_a$ are identified with the SU(2)-singlet
right-handed top quark with hypercharge 2/3, whereas the left-handed components
of $\psi_a$ denote the left-handed top quark, which transforms as a $I_3=1/2$
component of an (incomplete) SU(2) multiplet with isospin 1/2 and hypercharge
1/6. The complex scalar $\phi$ has the quantum numbers of the composite
operator $\bar\psi_L\psi_R$,  thus it transforms as the component of the Higgs
doublet whose VEV breaks the weak SU(2) and U(1) hypercharge down to
electromagnetism.

The set of operators to be taken into account in $S_{int}(\phi, \psi,\Lambda)$
has to be general enough for our purposes, and to respect the above-mentioned
symmetries (which can be expressed as linear combinations of the vector and
axial symmetries of the NJL model \cite{4}). Defining
 \be\label{2.8} D_p^n=\frac{d^4p_1}{(2\pi)^4}...\frac{d^4p_n}{(2\pi)^4}\delta^4
(\sum^n_{i=1}p_i)(2\pi)^4\ee the following decomposition of
$S_{int}(\phi,\psi,\Lambda)$ turns out to be sufficient:\vfill\newpage
\begin{eqnarray}\label{2.9} S_{int}(\phi,\psi,\Lambda)&= &\sum^\infty_{n=1}\int
D_p^{2n}F_{2n}(\Lambda;p_1,...,p_{2n})\phi^*(p_1)\phi
(p_2)...\phi^*(p_{2n-1})\phi(p_{2n})\nonumber\\ &+&\sum^\infty_{n=1}\int
D_p^{2n+1}\phi^*(p_1)\phi (p_2)...\phi^*(p_{2n-3})\phi(p_{2n-2})\nonumber\\
&&\cdot\bar\psi_a(p_{2n})G_{2n-1}(\Lambda;p_1,...,p_{2n+1})(\sigma(p_{2n-1})+
i\gamma_5\pi(p_{2n-1}))\psi_a(p_{2n+1})  \nonumber\\ &+&\sum^\infty_{n=1}\int
D_p^{2n+2}\phi^*(p_1)\phi (p_2)...\phi^*(p_{2n-1})\phi(p_{2n})\nonumber\\
&&\cdot\bar\psi_a(p_{2n+1})H_{2n}(\Lambda;p_1,...,p_{2n+2})
(-ip\llap/_{2n+2})\psi_a
(p_{2n+2})\end{eqnarray} (The index of the generalized couplings $F, G$ and $H$
corresponds to the number of scalars contained in the associated operator, and
the couplings $G$ and $H$ may be matrices with spinor indices.) This ansatz has
to be inserted into the flow equations (\ref{2.7}), which generally involve
contractions of both scalars with a propagator $P_\phi$ and fermions with a
propagator $P_\psi$. Then both sides have to be ordered according to their
field content, after which flow equations for the individual couplings
$F_{2n},G_{2n-1}$ and $H_{2n}$ are obtained. In the case of $F_{2n}$ these
equations have the following schematic form: \be\label{2.10} \partial_\Lambda
F_{2n}=N\partial_\Lambda P_\psi H_{2n}+\sum_m  \partial_\Lambda P_\phi
F_{2m}F_{2n-2m+2}+\partial_\Lambda P_\phi F_{2n+2}.\ee The factor $N$ in front
of the first term is due to $N$ pairs of fermions, which are contained in the
operator of the type $H$ and which  have to be contracted in order to turn it
into an operator of the type $F$. Let us, in order to study the $N\to\infty$
limit, perform the standard rescaling of fields and couplings:
\begin{eqnarray}\label{2.11} &&\phi\to\sqrt N\phi,\ F_{2n}\to N^{1-n}F_{2n},\
G_{2n-1}\to N^{1/2-n}\ G_{2n-1},\nonumber\\ &&H_{2n}\to
N^{-n}H_{2n}.\end{eqnarray} Now eq. (\ref{2.10}) becomes
 \be\label{2.12}  \partial_\Lambda F_{2n}=\partial_\Lambda P_\psi H_{2n}
+\sum_m \partial_\Lambda P_\phi F_{2m}F_{2n-2m+2}+N^{-1}\partial_\Lambda P_\phi
F_{2n+2}\ee and we see that the last term can be dropped. $N$-power counting
now also justifies the omittance of operators involving more than two fermionic
fields.

Next we have to remember that the couplings $F_{2n}$ describe connected Green
functions for $2n$ scalars including one-particle-reducible diagrams,  cf. the
general relation (\ref{2.6}). Finally we will be interested, however, in the
effective action or the generating functional of one-particle-irreducible
diagrams. In principle we have to perform a Legendre transformation in order to
get there; in terms of diagrams, however, we simply have to omit the
one-particle-reducible ones.

Generally it is not easy to modify the flow equation such that they describe
the  flow of the couplings of the effective action. In the present case,
however, this can be achieved by simply omitting the second term on the
right-hand side of equation (\ref{2.12}). Since the last term in (\ref{2.12})
has been dropped already, no self-contractions or tadpole-like diagrams
involving the $\phi$-fields are included any more, and it is true only under
these circumstances that the omittance of the second term corresponds only to
the omittance of  one-$\phi$-particle-reducible diagrams in the final
expression $S_{int}(\phi, \psi,\Lambda=0)$ and not more.

Thus, after omitting the second term in (\ref{2.12}), we have not made any
approximation, but changed the physical interpretation of the couplings
$F_{2n}$, which now describe the coefficients of the effective action for the
fields $\phi$.

Generally a local effective action can be expanded in powers of derivatives or,
in momentum space, in powers of $p$. Subsequently we will be interested in the
effective potential, which corresponds to the zeroth order in such an
expansion, and the wave function normalization of the field $\phi$, for which
one needs the second order of  the effective action in $p$. From the structure
of the flow equation one then finds that it is sufficient to know all the
couplings $F$, $G$, and $H$ only up to second order in an expansion in the
momenta associated  with the fields $\phi$, since no $\phi$ contractions
appear any more.

To start with, however, we will concentrate on the effective potential and thus
deal with the couplings $F$, $G$, and $H$ at vanishing bosonic momenta only.
Therefore we define \begin{eqnarray}\label{2.13} &&F_{2n}(\Lambda;0,..,0)\equiv
F_{2n}^{(0)},\nonumber\\ &&G_{2n-1}(\Lambda;0,..,0,p_{2n}=p,p_{2n+1}=-p)\equiv
G^{(0)}_{2n-1}(p),\\ &&H_{2n}(\Lambda;0,..,0,p_{2n+1}=p, p_{2n+2}=-p)\equiv
H^{(0)}_{2n}(p).\nonumber \end{eqnarray}

In the case of the restricted couplings of the form (\ref{2.13}) the ansatz
(\ref{2.9}) for $S_{int}$ can most conveniently be expressed in terms of fields
$\phi$ of the form \be\label{2.14} \phi(p)=\delta^4(p)\hat\phi\ee and
\begin{eqnarray}\label{2.15} &&\hat
F(\Lambda;\hat\phi)=\sum^\infty_{n=1}F_{2n}^{(0)}(|\hat\phi|^2)^n\nonumber\\
&&\hat
G(\Lambda;\hat\phi,p)=\sum^\infty_{n=1}G^{(0)}_{2n-1}(p)(|\hat\phi|^2)^{n-1} \\
&&\hat
H(\Lambda;\hat\phi,p)=\sum^\infty_{n=1}H^{(0)}_{2n}(p)(|\hat\phi|^2)^{n}.
\nonumber\end{eqnarray} Now (\ref{2.9}) becomes \begin{eqnarray}\label{2.16}
&&S_{int}(\phi,\psi,\Lambda)=\hat F(\Lambda;\hat\phi)\delta^4(0)(2\pi)^4+
\int\frac{d^4p}{(2\pi)^4}\\ &&\Bigl\lbrace\bar\psi_a(p)\hat
G(\Lambda;\hat\phi,p)(\hat\sigma+i\gamma_5
\hat\pi)\psi_a(-p)+\bar\psi_a(p)\hat
H(\Lambda;\hat\phi,p)(-ip\llap/)\psi_a(-p)
\Bigr\rbrace.\nonumber\end{eqnarray} As usual $(2\pi)^4\delta^4(0)$ is to be
interpreted as the space-time volume. Note that $\hat F(\Lambda;\hat\phi)$
corresponds to nothing but the $\Lambda$-dependent effective potential.

The flow equations have the form of eq. (\ref{2.7}) where, after $1/N$ counting
and the previous discussion, only fermionic contractions occur. The regularized
fermionic propagator has the form
 \be
\label{2.17}P_{\psi,\Lambda}^{\Lambda_0}(p^2)=-\frac{if(\Lambda,p)}{p\llap/}\ee
with $f(\Lambda,p)$ as in eq. (\ref{2.2}). Then, after inserting the ansatz
(\ref{2.9}) with (\ref{2.13}) into (\ref{2.7}), performing the necessary traces
over $\gamma$ matrices and ordering the resulting equation according to their
field content, one finds the following flow equations for the couplings of
(\ref{2.13}): \begin{eqnarray}\label{2.18} &&\partial_\Lambda
F^{(0)}_{2n}=-4\int\frac{d^4p}{(2\pi)^4}\partial_\Lambda f(\Lambda
,p)H^{(0)}_{2n}(p),\nonumber\\ &&\partial_\Lambda
G^{(0)}_{2n-1}(p)=-2\sum^n_{m=1}\partial_\Lambda f(\Lambda,p)
G^{(0)}_{2m-1}(p)H^{(0)}_{2n-2m}(p),\nonumber\\ &&\partial_\Lambda
H^{(0)}_{2n}(p)=-\sum^n_{m=0}\partial_\Lambda f(\Lambda,p)
H^{(0)}_{2m}(p)H^{(0)}_{2n-2m}(p)\nonumber\\
&&\qquad\qquad+\sum^n_{m=1}\frac{\partial_\Lambda f(\Lambda,p)}{p^2}
G^{(0)}_{2m-1}(p)G^{(0)}_{2n-2m+1}(p).\end{eqnarray} This infinite set of
coupled differential equations can actually be handled by use of the quantities
defined in (\ref{2.15}). In terms of $\hat F,\hat G$ and $\hat H$ and after a
rearrangement of indices the flow equations (\ref{2.18}) simply become
\begin{eqnarray}\label{2.19} &&\partial_\Lambda\hat
F(\Lambda;\hat\phi)=-4\int\frac{d^4p}{(2\pi)^4}\partial_\Lambda
f(\Lambda,p)\hat H(\Lambda;\hat\phi,p),\nonumber\\ &&\partial_\Lambda\hat
G(\Lambda;\hat\phi,p)=-2~\partial_\Lambda f(\Lambda,p)\hat
G(\Lambda;\hat\phi,p)\hat H(\Lambda;\hat\phi,p),\nonumber\\
&&\partial_\Lambda\hat H(\Lambda;\hat\phi,p)=-\partial_\Lambda
f(\Lambda,p)\left\lbrack{\hat H}^2(\Lambda;\hat\phi,p)-\hat
G^2(\Lambda;\hat\phi,p) \frac{|\hat\phi|^2}{p^2}\right\rbrack.\end{eqnarray}

After an expansion of eqs. (2.19) in powers of $\hat\phi$ one recovers that
they coincide with the infinite system (\ref{2.18}). On the other hand, the
solutions to eqs. (\ref{2.19}) with arbitrary boundary conditions at
$\Lambda=\Lambda_0$, \begin{eqnarray}\label{2.20} &&\hat F(\Lambda_0;\hat
\phi)\equiv\hat F_0(\hat\phi),\nonumber\\  &&\hat G(\Lambda_0;\hat
\phi,p)\equiv\hat G_0(\hat\phi,p),\nonumber\\  &&\hat H(\Lambda_0;\hat
\phi,p)\equiv\hat H_0(\hat\phi,p),\end{eqnarray} are easily
found:\vfill\newpage \begin{eqnarray}\label{2.21} &&\hat
F(\Lambda;\hat\phi)=-2\int\frac{d^4p}{(2\pi)^4}\ln[1+2f(\Lambda,p)\hat
H_0+f^2(\Lambda,p)(\hat H_0^2+\hat G^2_0\frac{|\hat\phi |^2}{p^2})] +\hat
F_0,\nonumber\\ &&\hat G(\Lambda;\hat\phi,p)=\frac{\hat
G_0}{(1+f(\Lambda,p)\hat H_0)^2+f^2(\Lambda, p)\hat
G^2_0|\hat\phi|^2/p^2}\nonumber\\ &&\hat H(\Lambda;\hat\phi,p)=\frac{\hat
H_0+f(\Lambda,p)(\hat H_0^2+\hat G_0^2|\hat\phi |^2/p^2)}{(1+f(\Lambda,p)\hat
H_0)^2+f^2(\Lambda ,p)\hat G_0^2|\hat\phi|^2/p^2}.\end{eqnarray}

Let us recall the physical interpretation of the quantities  $\hat
F(\Lambda;\hat\phi),\hat G(\Lambda;\hat\phi,p)$ and $\hat H(\Lambda;\hat\phi,
p)$. Whereas $\hat F$ is related to the effective potential for $\hat\phi$ (up
to a factor $N$ implied by the rescaling (\ref{2.11})), $\hat G$ and $\hat H$
appear in the $\psi$-dependent part of $S_{int}$, see (\ref{2.16}). In order to
relate $\hat G$ and $\hat H$ to the full fermionic propagator $S_F$ in the
background $\hat\phi$, we have to take into account that $S_{int}$ is  related
to the generating functional of connected Green functions, cf. (\ref{2.6}). In
the  present case, where $S_{int}$ is just quadratic in the fermionic fields,
the analog of (\ref{2.6}) reads \be\label{2.22}
S_{int}(\psi=P_{\psi,0}^{~\Lambda_0}~\eta
,\Lambda=0)=\int\frac{d^4p}{(2\pi)^4}\left\lbrace  -\bar\eta S_F\eta+\bar\eta
P_{\psi,0}^{~\Lambda_0}\eta\right\rbrace.\ee Here $P_{\psi,0}^{~\Lambda_0}$ is
the free cutoff propagator as in (\ref{2.17}), and the full propagator $S_F$
can now be determined by inverting (\ref{2.22}) and using (\ref{2.16}) with
(\ref{2.21}): \be\label{2.23}
S_F(\hat\phi,p)=\frac{f(0,p)}{ip\llap/(1+f(0,p)\hat H_0(\hat\phi,p))+f(0,p)\hat
G_0(\hat\phi,p) (\hat\sigma+\gamma_5\hat\pi)}.\ee

For dimensional reasons the boundary conditions $\hat F_0$, $\hat G_0$ and
$\hat H_0$ of eq. (\ref{2.20}) have to be of the following form:
\begin{eqnarray}\label{2.24} &&\hat
F_0(\hat\phi)=\alpha\Lambda^4_0+\frac{\mu}{2N}\Lambda^2_0\hat\phi^2+
\frac{\lambda_0}
{2N}\hat\phi^4+O\left(\frac{\hat\phi^2}{\Lambda^2_0}\right),\nonumber\\ &&\hat
G_0(\hat\phi,p)=g\left(\frac{p^2}{\Lambda^2_0}\right)+O
\left(\frac{\hat\phi^2}{\Lambda^2_0}\right),\nonumber\\ &&\hat
H_0(\hat\phi,p)=h\left(\frac{p^2}{\Lambda^2_0}\right)+O
\left(\frac{\hat\phi^2}{\Lambda^2_0}\right)\end{eqnarray} with arbitrary
dimensionless parameters $\alpha$, $\mu$ and $\lambda_0$ and  arbitrary
functions g and h.  Here we assumed that the bare action is local (i.e. $\hat
G_0$ and $\hat H_0$  are finite for $p^2\to0$) and involves only positive
powers of the field $\phi$.

Subsequently we will be interested in the situation where the scale of physics,
represented by the Higgs VEV $\hat\phi$, is much smaller than the scale
$\Lambda_0$ of the bare action. This allows us to neglect the last terms in
(\ref{2.24}), and the effective potential given by $N\hat
F(\Lambda=0;\hat\phi)$ can be written as
 \be\label{2.25}
V_{eff}=\frac{\mu}{2}\Lambda^2_0\hat\phi^2+\frac{\lambda_0}{2}
\hat\phi^4-\frac{N}{8\pi^2}\int^2_0xdx\ln\left[1+
\frac{|\hat\phi|^2}{x\Lambda^2_0}k(x)\right]\ee with
 \be\label{2.26} k(x)=\frac{f^2(\Lambda=0,p^2=x\Lambda^2_0)g^2(x)}
{(1+f(\Lambda=0,p^2=x\Lambda^2_0)\cdot h(x))^2}.\ee We have dropped a
$\hat\phi$-independent constant piece, and the upper limit 2 of the $x$
integration is due to the fact that $f(x)$ and hence  $k(x)$ vanish for $x>2$
according to eqs. (\ref{2.2}) and (\ref{2.3}).

Note that, before putting $\Lambda=0$ in (\ref{2.21}), $N\hat F(\Lambda;
\hat\phi)$ interpolates continuously between the bare potential  $N\hat F_0$
and the full effective potential  (\ref{2.25}). In the limit of a sharp cutoff,
$\tilde\theta\to\theta$ in (\ref{2.2}), and for the particularly simple
boundary conditions $g(x)=const.=g$ and $h(x)=0$ in (\ref{2.24}) the
``flowing'' potential $V_{eff}(\Lambda,\hat\phi)$ becomes
\begin{eqnarray}\label{2.27}
&&V_{eff}(\Lambda,\hat\phi)=\frac{\mu}{2}\Lambda^2_0\hat\phi^2+\frac{\lambda_0}
{2}
\hat\phi^4+\frac{N}{16\pi^2}\Bigl\lbrace g^2\hat\phi^2(\Lambda^2-\Lambda^2_0)
\nonumber\\
&&+g^4\hat\phi^4\ln\left(\frac{\Lambda^2_0+g^2\hat\phi^2}
{\Lambda^2+g^2\hat\phi^2}
\right)+\Lambda^4\ln\left(1+\frac{g^2\hat\phi^2}{\Lambda^2}\right)-\Lambda_0^4
\ln\left(1+\frac{g^2\hat\phi^2}{\Lambda^2_0}\right)\Bigr\rbrace.\end{eqnarray}
(A similar formula for different models can be found in \cite{18}.)

One finds that $V_{eff}(\Lambda,\hat\phi)$ develops a nontrivial minimum in
$\hat\phi$ as soon as the infrared  scale $\Lambda$ satisfies \be\label{2.28}
\Lambda^2\leq\Lambda^2_0\left(1-\frac{4\pi^2\mu}{Ng^2}\right).\ee

In the case of general  boundary conditions the formula (\ref{2.25}) for
$V_{eff}$ can still be evaluated, neglecting powers of $\hat\phi/\Lambda_0$.
After some algebra one obtains \be\label{2.29}
V_{eff}=\hat\phi^2\frac{\Lambda^2_0}{2}\left(\mu-\frac{Na}{4\pi^2}\right)+
\frac{\hat\phi^4}{2}\left\lbrace\lambda_0+\frac{N}{8\pi^2}\left[
b_0+b_1\ln\left(
\frac{\Lambda^2_0}{\hat\phi^2k(0)}\right)\right]\right\rbrace\ee with
\begin{eqnarray}\label{2.30} &&a=\int^2_0dx\ k(x),\nonumber\\
&&b_0=\int^2_0\frac{dx}{x}(k^2(x)-k^2(0))+k^2(0)(\frac{1}{2}+\ln(2)),
\nonumber\\
&&b_1=k^2(0).\end{eqnarray} Note that $V_{eff}$, according to eq.
(\ref{2.29}), has the same form as in the renormalization group-improved
standard model to leading order in $1/N_c$ (where $g_t$ denotes the top quark
Yukawa coupling): \be\label{2.31}
V^{SM}_{eff}=\frac{m^2}{2}\hat\phi^2+\frac{\hat\phi^4}{2}
\left\lbrace\lambda(\mu)+\frac{N_c
g_t^4}{8\pi^2}\ln\left(\frac{\mu^2}{\hat\phi^2} \right)\right\rbrace\ee This
correspondence has been achieved for virtually arbitrary boundary conditions
at $\Lambda_0$, we only used $\hat\phi/\Lambda_0\ll1$. The explicit relations
between the parameters are \be\label{2.32}
m^2=\Lambda^2_0\left(\mu-\frac{Na}{4\pi^2}\right),\ee \be\label{2.33}
\lambda(\mu)=\lambda_0+\frac{N}{8\pi^2}\left[
b_0+b_1\ln\left(\frac{\Lambda_0^2} {\mu^2k(0)}\right)\right],\ee and, for the
top quark Yukawa coupling $g_t$, \be\label{2.34} g^2_t=k(0).\ee Of course fine
tuning between the parameters $\mu$ and $a$, which are specified by the
boundary conditions at $\Lambda_0$, is required in order to achieve
$\hat\phi\ll\Lambda_0$ at the minimum of $V_{eff}$ of eq. (\ref{2.29}).
Whereas the fine tuning problem is automatically obvious within the present
framework, there is no conceptual difficulty in assuming
$\hat\phi\ll\Lambda_0$ and studying the consequences.

In this limit and for momenta $p$ with $p^2/\Lambda^2_0\ll1$ the fermion
propagator $S_F$ of (\ref{2.23}) simplifies as well; with
$f(0,p\ll\Lambda_0)=1$ and taking $\hat\pi=<\pi>=0$ by definition one finds
\be\label{2.35}
(1+h(0))S_F(\hat\phi,p)=\frac{1}{ip\llap/+\frac{g(0)\hat\sigma}{1+h(0)}}.\ee
The constant factor $(1+h(0))$ on the left-hand side can be absorbed by a
rescaling of the fields $\psi_a$, and the right-hand side is the top quark
propagator of the standard model with $g_t=g(0)/(1+h(0))$ or $g_t^2=k(0)$ in
agreement with eq. (\ref{2.34}).

In order to relate the Higgs VEV $\hat\phi=\hat\sigma$ to the gauge boson
masses, and the second derivative of the effective potential to the mass of
the physical Higgs fields $\sigma$ (with the mass defined by the inverse
propagator at zero momentum), the wave function normalization $Z_{eff}$ of the
real component $\sigma$ of the complex field $\phi$ has to be known as well.
$Z_{eff}$ appears in the effective action in the form
 \be\label{2.36} \frac{1}{2}Z_{eff}(\hat\phi)\int
d^4p\sigma(p)p^2\sigma(-p).\ee

This term can be obtained from the couplings $F_{2n}$ of (\ref{2.9}), once they
are expanded to second order in the momenta $p_i$ and to second order in the
deviation of the field $\sigma$ from its VEV $\hat\phi$. From the structure of
the flow equations, eqs. (\ref{2.18}), it follows that the knowledge of
$F_{2n}$, expanded to second order in the momenta $p_i$, requires also the
knowledge of $H_{2n}$ and $G_{2n-1}$ expanded to second order in the momenta
associated with the scalars $\phi$. Unfortunately the number of possible
Lorentz invariants which can be formed out of two ``bosonic'' momenta $p_i$
and $p_j$, a momentum $p$ of a fermion (e.g. the one associated with the
fields $\psi_a$ in eqs. (\ref{2.9})) and $\gamma$ matrices is quite large.
Thus, instead of eq. (\ref{2.13}), we now have the following expansion
(asssuming the couplings to be symmetric in the bosonic momenta):
\begin{eqnarray}\label{2.37} F_{2n}&=&F_{2n}^{(0)}+p^2_iF_{2n}^{(1)}+p_i\cdot
p_j F^{(2)}_{2n}+O(p^4_i)\nonumber\\ G_{2n-1}&=&G^{(0)}_{2n-1}+p_i\cdot
pG^{(1)}_{2n-1}+p\llap/_ip\llap/G^{(2)}_{2n-1}+p^2_i G_{2n-1}^{(3)}\nonumber\\
&&+p_i\cdot pp\llap/_ip\llap/G^{(4)}_{2n-1}+(p_i\cdot p)^2
G^{(5)}_{2n-1}+p_i\cdot p_j G^{(6)}_{2n-1}\nonumber\\ &&+p_i\cdot
pp\llap/_jp\llap/ G^{(7)}_{2n-1}+(p_i\cdot p)(p_j\cdot p)
G^{(8)}_{2n-1}+O(p^3_i)\nonumber\\ H_{2n-1}&=&H^{(0)}_{2n}+p_i\cdot
pH^{(1)}_{2n}+p\llap/_ip\llap/H^{(2)}_{2n}+p^2_i H_{2n}^{(3)}\nonumber\\
&&+p_i\cdot pp\llap/_ip\llap/H^{(4)}_{2n}+(p_i\cdot p)^2 H^{(5)}_{2n}+p_i\cdot
p_j H^{(6)}_{2n}\nonumber\\ &&+p_i\cdot pp\llap/_jp\llap/
H^{(7)}_{2n}+(p_i\cdot p)(p_j\cdot p) H^{(8)}_{2n}+O(p^3_i)\end{eqnarray}

The couplings $F^{(i)},G^{(i)}$ and $H^{(i)}$ do not depend on any ``bosonic''
momenta by definition, but $G^{(i)}$ and $H^{(i)}$ depend on the ``fermionic''
momentum $p$. After some combinatorics one finds that the wave function
normalization $Z_{eff}$ of (\ref{2.36}) is related to the  couplings
$F_{2n}^{(1)}$ and $F_{2n}^{(2)}$ as \be\label{2.38}
Z_{eff}(|\hat\phi|)=\sum^\infty_{n=1}(|\hat\phi|^2)^{n-1}
\lbrace(2n-1)F_{2n}^{(1)}-F_{2n}^{(2)}\rbrace.\ee

It is straightforward, though not very illuminating, to derive the system of
flow equations for the couplings $F^{(i)},G^{(i)}$ and $H^{(i)}$ and to solve
it along the line of eqs. (\ref{2.18}), (\ref{2.15}), and (\ref{2.19}). More
details will be presented in a forthcoming publication \cite{19}, here we just
present the result for $Z_{eff}$ for the following class of boundary
conditions at $\Lambda_0$:  Defining, in analogy to (\ref{2.18}),
\be\label{2.39} \hat
F^{(i)}(\Lambda;\hat\phi)=\sum^\infty_{n=1}F_{2n}^{(i)}(|\hat\phi|^2)^n \ee
and $\hat G^{(i)}$ and $\hat H^{(i)}$ correspondingly, we take at
$\Lambda=\Lambda_0$: \begin{eqnarray}\label{2.40} &&\hat
F^{(i)}(\Lambda_0;\hat \phi)\equiv\hat F^{(i)}_0(\hat\phi),\nonumber\\  &&\hat
G^{(0)}(\Lambda_0;\hat \phi,p)\equiv\hat G^{(0)}_0(\hat\phi,p)=
g\left(\frac{p^2}{\Lambda^2_0}\right)+O
\left(\frac{\hat\phi^2}{\Lambda^2_0}\right),\nonumber\\ &&\hat
G^{(i)}(\Lambda_0;\hat \phi,p)\equiv\hat G^{(i)}_0(\hat\phi,p)=
g_i\left(\frac{p^2}{\Lambda^2_0}\right)+O
\left(\frac{\hat\phi^2}{\Lambda^2_0}\right)\quad for \quad i = 1,3,5,
\nonumber\\ &&\hat G^{(i)}(\Lambda_0;\hat \phi,p)\equiv\hat
G^{(i)}_0(\hat\phi,p)=0 \qquad for \quad i \neq 0,1,3,5,\nonumber\\ &&\hat
H^{(0)}(\Lambda_0;\hat \phi,p)\equiv\hat H^{(0)}_0(\hat\phi,p)=
h\left(\frac{p^2}{\Lambda^2_0}\right)+O
\left(\frac{\hat\phi^2}{\Lambda^2_0}\right),\nonumber\\ &&\hat
H^{(i)}(\Lambda_0;\hat \phi,p)\equiv\hat H^{(i)}_0(\hat\phi,p)=0 \qquad for
\quad i \neq 0. \end{eqnarray} With $f \equiv f(0,p^2/\Lambda_0^2)$ and primes
denoting $\partial/\partial p^2$ we  obtain for $Z_{eff}$ in the limit
$|\hat\phi|\ll\Lambda_0$: \bea Z_{eff}(|\hat\phi|)= & \frac{N}{4\pi^2}
\int_0^{2\Lambda_0^2}dp^2 \Biggl\lbrace
\Biggl[\frac{3p^4}{2((1+h)^2p^2+g^2|\hat \phi|^2)^3} -\frac{p^6(1+h)^2}
{((1+h)^2p^2+g^2|\hat \phi|^2)^4} \Biggr] g^2f^2(1+h)^4\nonumber\\
&+\frac{2gg_3f^2}{(1+h)^2}+\frac{g^2ff'}{2(1+h)^2}-\frac{g^2f^2h'}{2(1+h)^3}+
\frac{gg_1f^2}{2(1+h)^2}\nonumber\\
&+p^2\Biggl(-\frac{gg_5f^2}{2(1+h)^2}-\frac{g_1^2f^2}{4(1+h)^2}-
\frac{g^2ff''}{2(1+h)^2}+\frac{gg_1ff'}{(1+h)^2}\nonumber\\
&-\frac{gg_1f^2h'}{(1+h)^3}+\frac{g^2ff'h'}{(1+h)^3}-
\frac{g^2f^2h'^2}{(1+h)^4} \Biggr)\Biggr]\Biggr\rbrace +
Z_{eff}(|\hat\phi|)_{0}. \label{2.41} \eea Here $Z_{eff}(|\hat\phi|)_{0}$ is
determined by the $\hat F^{(i)}_0$ of  (\ref{2.40}) in analogy to
(\ref{2.38}). The first term on the right hand side contains a logarithmically
divergent  part leading to \be\label{2.42}
Z_{eff}(|\hat\phi|)=\frac{N}{8\pi^2}
k(0)\ln\left(\frac{\Lambda^2_0}{|\hat\phi|^2}\right)\ +\ finite \ee with
$k(0)$ as in (\ref{2.26}). Note that the finite part contains a term
involving  $f''$ which leads to a divergence in the limit where $\tilde\theta$
approaches the sharp $\theta$ function. This has nothing to do with the use of
the flow equations themselves, but with the use of cutoff propagators as
(\ref{2.1}) or (\ref{2.17}). As discussed in the appendix of \cite{16}, the
effect can already be seen by computing $Z_{eff}$ with standard methods using
cutoff propagators.

It is thus clear that the physical predictions of the model depend not only on
the bare action in the form of the functions $g_i$ and $h$, but also on the
form of the cutoff. This is quite plausible, though, since at least the
present way of implementing the cutoff can be considered as a regularization
through higher  derivatives and thus the addition of higher-dimensional
operators to the action, which has similar physical effects as the addition of
higher dimensional operators in the form of the functions $g_i$ and $h$.

\section{The Generalized NJL Models} \setcounter{equation}{0}

In the NJL model the only field, to start with, is a Dirac spinor $\psi_a$. The
action is invariant under vector and axial vector symmetries, and contains a
pointlike four-Fermi interaction \cite{4}. It has been proposed to apply the
model to the Higgs-top sector of the standard model \cite{3}: Then the Higgs
scalar can be interpreted as a top-antitop bound state, and predictions for the
Higgs and top quark masses arise. According to \cite{5}-\cite{7}, however,
these predictions get lost if the model is generalized by the addition of
higher derivative couplings (which have to be expected if the model is
considered as an effective low energy theory of something else).

The class of generalized NJL models we will consider is defined by a bare
action of the form \begin{eqnarray}\label{3.1} \tilde
S^0&=&\int\frac{d^4p}{(2\pi)^4}(1+\tilde h(p))\bar\psi_a(p)(-ip\llap/)
\psi_a(p)\nonumber\\ &&-\int\frac{d^4p}{(2\pi)^4} \frac{d^4p_1}{(2\pi)^4}
\frac{d^4p_2}{(2\pi)^4}  \frac{N\tilde g(p,p_1)\tilde g(-p,p_2)}{2\Lambda_0^2}
\Bigl\lbrace\bar\psi_a(p)\psi_a(-p-p_1)\bar\psi_b(p)\psi_b(p-p_2)\nonumber\\
&&-\bar\psi_a(p)\gamma_5\psi_a(-p-p_1)
\bar\psi_b(p)\gamma_5\bar\psi_b(p-p_2)\Bigr
\rbrace\end{eqnarray} where, on dimensional grounds, the functions $\tilde h$
and $\tilde g$ depend on $p_i$ only via the dimensionless ratio
$p_i/\Lambda_0$. The functions $\tilde h$ and $\tilde g$ generate, once
expanded in powers of the momenta, derivative couplings which are more general
than the ones considered in \cite{5}, \cite{6}.

The action (\ref{3.1}) can be rewritten by the introduction of the complex
scalar field $\phi=\sigma+i\pi$. The action \begin{eqnarray}\label{3.2}
S^0&=&\int\frac{d^4p}{(2\pi)^4}(1+\tilde
h(p))\bar\psi_a(-p)(-ip\llap/)\psi_a(p) \nonumber\\
&&+\int\frac{d^4p}{(2\pi)^4}\frac{\Lambda^2_0}{2N} \phi^*(p)\phi(-p)\nonumber\\
&&+\int\frac{d^4p}{(2\pi)^4}\frac{d^4p_1}{(2\pi)^4}\tilde g(p,p_1)
\bar\psi_a(p)(\sigma(p_1)+i\pi(p_1)\gamma_5)\psi_a(-p-p_1)\end{eqnarray} is
easily seen to coincide with the one of eq. (3.1) once the real scalars
$\sigma$ and $\pi$ are integrated out or replaced by their classical equations
of motion.

On the other hand, the action (3.2) belongs to the class of actions treated in
sect. 2, cf. eq. (\ref{2.9}). Note that the action (\ref{3.2}) is to be
considered as a bare action, which has to be used as boundary condition for the
running action $S_{int}(\psi,\phi,\Lambda)$ at $\Lambda=\Lambda_0$. The general
boundary conditions for those couplings, which are relevant for the effective
potential and the full fermionic propagator for $\hat\phi\ll\Lambda_0$, have
previously been formulated in eqs. (\ref{2.24}). They involved two parameters
$\mu$ and $\lambda_0$ and two arbitrary functions $g(p)$ and $h(p)$.

In order to coincide with the bare action (\ref{3.2}) of a generalized NJL
model, these boundary conditions have to be specified as
\begin{eqnarray}\label{3.3} \mu&=&1,\nonumber\\ \lambda_0&=&0,\nonumber\\
g(p)&=&\tilde g(p,0),\nonumber\\ h(p)&=&\tilde h(p).\end{eqnarray} Since we
have already computed the effective potential and the fermionic propagator of
the low energy theory in the most general case, we can investigate whether the
restrictions (\ref{3.3}) for $\mu$ and $\lambda_0$ result in constraints on the
three parameters $m^2,\lambda(\mu)$ and $g_t$ in (\ref{2.31}).  From eqs.
(\ref{2.32}), (\ref{2.33}), and (\ref{2.34}) together with (\ref{2.30}) and
(\ref{2.26}), however, it is easily seen that for general functions $\tilde g$
and $\tilde h$ no constraints emerge. Actually, even if the function $\tilde g$
and $\tilde h$ would be fixed (e.g. by $\tilde g=const.=g$ and $\tilde h=0$)
predictions arise only if the detailed form of the cutoff, i.e. the function
$f(\Lambda,p)$ or the $\tilde\theta$ function in eq. (\ref{2.2}), is fixed as
well.

The same result emerges in the case of the wave function normalization
$Z_{eff}$ of  eq. (\ref{2.39}), which is required in order to relate the
parameters of $V_{eff}$ to physical quantities. The bare action (\ref{3.2})
corresponds to the boundary condition
 \be\label{3.4} Z_{eff}(\hat\phi,\Lambda=\Lambda_0)\equiv Z_{eff0}=0.\ee Again
this constraint is not sufficient to restrict $Z_{eff}$ of eq. (\ref{2.39})
unless the functions $f,g$ and $h$ are further specified.

Thus we have rederived the results of \cite{5}-\cite{7}, though in an even more
general framework and by a different method, that generalized NJL models are
not restrictive enough to allow for predictions of the parameters of the
effective low energy theory. On the other hand we can easily check that in the
limit $\ln(\Lambda_0^2/\hat\phi^2)\gg1$ the relation $m_\sigma=2m_t$ is
obtained independently of the boundary conditions \cite{20},\cite{21}.

\section{Discussion and Outlook} The flow equations have allowed us to relate
arbitrary bare actions for the Higgs top sector to effective low energy
theories, which have as many free parameters as a general renormalizable model,
as it should be \cite{10}. We could  investigate whether the constraints
imposed on the bare action by requiring it to be of the  generalized NJL-type
(\ref{3.2}) lead to relations among the low energy parameters. The negative
answer was known before \cite{5}-\cite{7}, just our ansatz (\ref{3.2}) is even
more general than the one of \cite{5},\cite{6}. Also we have verified the
additional dependence of the low energy parameters on the form of the cutoff
\cite{7}. This affects the original version of the NJL  model even if the
couplings of the bare action are fixed.

The emphasis of the present paper is, however, on the method by which our
results have been obtained. The flow equations make the intuitive picture of
scale-dependent effective actions manifest. Here we have studied an example,
where the $1/N$ expansion allows us to find non-perturbative solutions to these
equations (which have been presented in detail, however, only for the couplings
relevant for the effective potential). With these solutions we can study the
convergence and irreversibility of the flow of actions towards the infrared;
this phenomenon essentially keeps us from gaining knowledge about a possible
underlying theory valid at some large scale $\Lambda_0$, unless we perform
measurements with a precision of $\sim O(Q^2/\Lambda^2_0)$ where $Q^2$ denotes
the energy scale of the process.

The flow equations, furthermore, can possibly be used to study more complicated
strongly interacting systems. Then the infinite set of coupled differential
equations will necessarily have to be approximated, typically by restricting
the set of operators in $S_{int}$. It will thus be of great help that a
solvable  system of these equations exists; this can then be used to test
approximation schemes. To our knowledge the system present here is the first
one satisfying these requirements.

\end{document}